\documentclass[12pt]{iopart}
\usepackage[dvips]{graphicx}
\usepackage{color}
\usepackage{hyperref}

\eqnobysec

\begin{document}


\letter{Stellar parallax in the Neo-Tychonian planetary system}

\author{Luka Popov \vspace{2mm}}

\address{University of Zagreb, Department of Physics,
Bijeni\v cka cesta 32, Zagreb, Croatia}

\ead{lpopov@phy.hr}

\begin{abstract} \noindent
  The recent paper published in European Journal of Physics \cite{popov} aimed
  to demonstrate the kinematical and dynamical equivalence of heliocentric
  and geocentric systems. The work is performed in the Neo-Tychonian system,
  with key assumption that orbits of distant masses around the Earth are
  synchronized with the Sun's orbit. Motion of Sun and Mars have been analysed,
  and the conclusion was reached that the very fact of the accelerated motion
  of the Universe as a whole produces the so-called ``pseudo-potential'' that
not
  only explains the origin of the pseudo-forces, but also the very 
  motion of the celestial bodies as seen from the static Earth. After the
  paper was published, the question was raised if that same potential can
  explain the motion of the distant stars that are not affected by the 
  Sun's gravity (unlike Mars), and if it can be used to reproduce the
observation
  of the stellar parallax. The answer is found to be positive.
\end{abstract}



\submitto{\EJP}

 \maketitle

\setcounter{equation}{0}
\section{Introduction}\label{intro}

The well-known effect of stellar parallax can be explained in two ways. The
first and most common one is in the heliocentric system, in which the Sun and
the observed stars are approximately considered to be at rest. While the Earth
moves around the Sun, its position relative to the stars changes, and that 
results with the effect of motion of the near stars \cite{carrol}. The parallax 
is observed using the more distant stars in the background.

Second way to explain stellar parallax is by saying that the apparent movement
of the stars is in fact the real motion in the pseudo-potential that is,
according to
Mach's principle \cite{barbour}, generated by the very fact of the simultaneous
accelerated motion of all the bodies in the Universe, including the distant
stars.

\begin{figure}[t]
  \centering
  \includegraphics[scale=0.3]{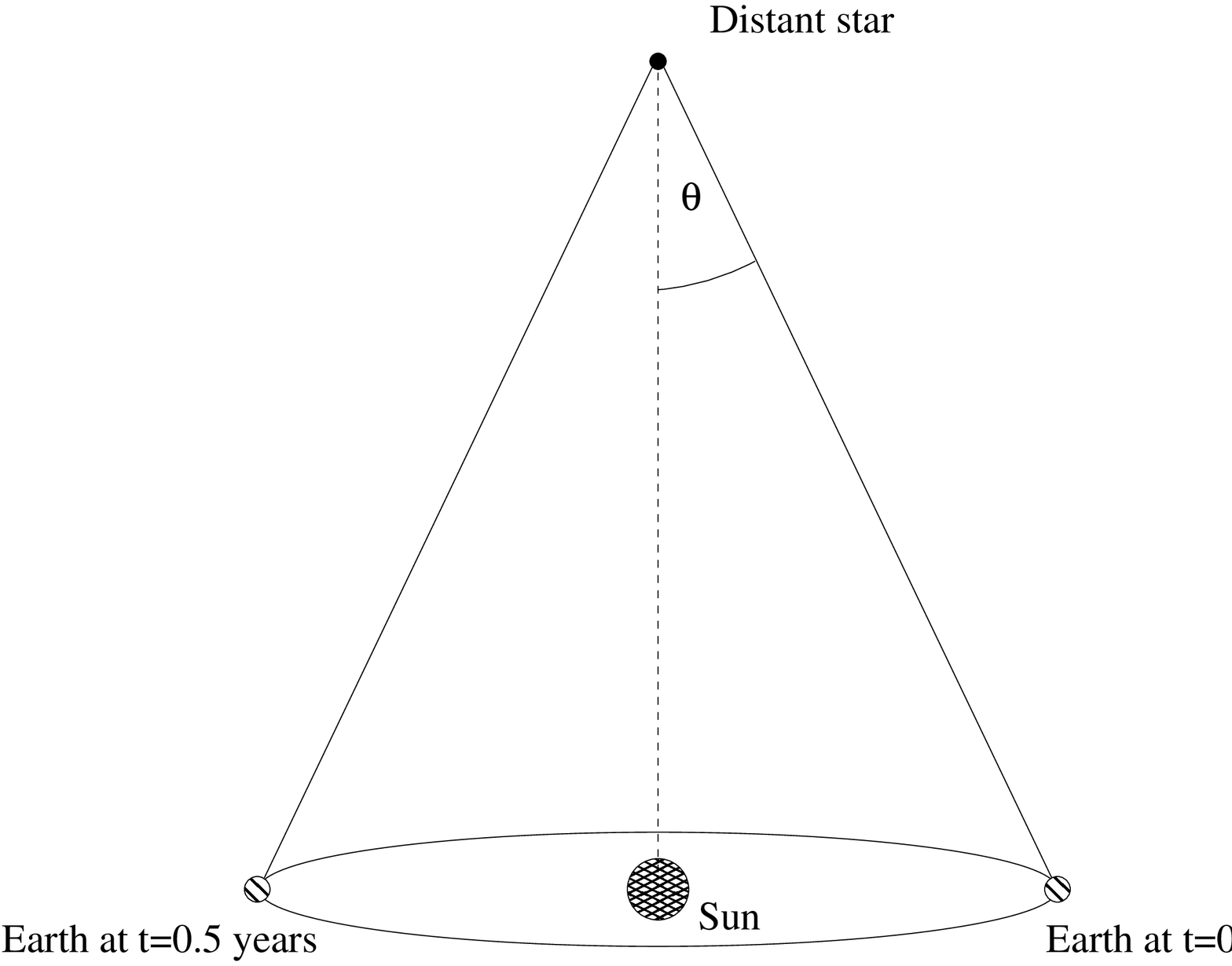} $\qquad$
  \includegraphics[scale=0.3]{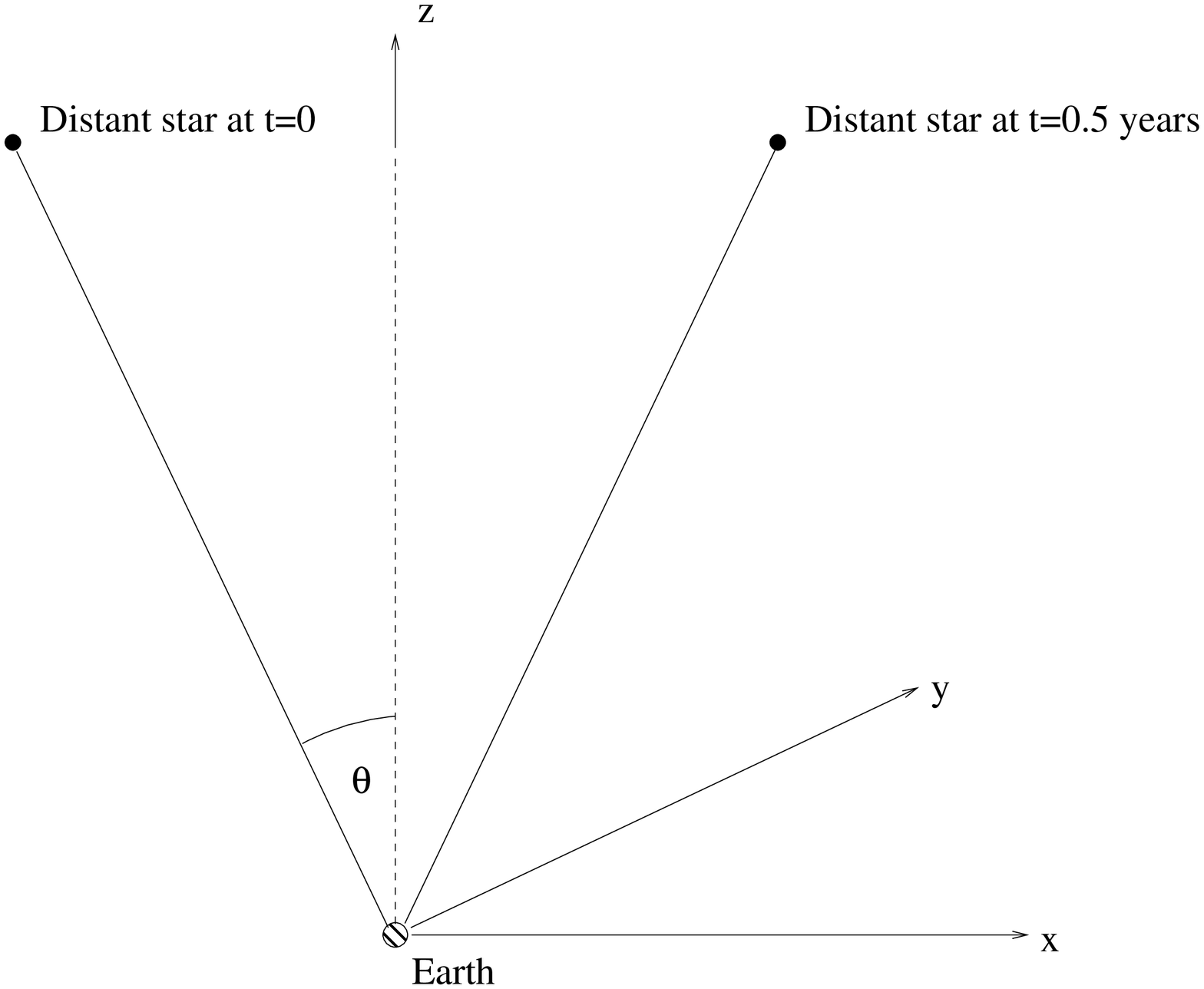}
  \caption{\label{fig1} Illustrations of the stellar parallax in the
heliocentric (left pannel) versus geocentric (right pannel) frames of
reference.}
\end{figure}

The comparison between two approaches is given in the Figure \ref{fig1}, with
the appropriate choice of coordinate axes that will be used in the calculation
which follows.

\section{Motion of Proxima Centauri in the Earth's pseudo-potential}

Now in order to demonstrate how one can arrive to the correct prediction of
the stellar parallax in the Neo-Tychonian system, we will calculate the
trajectory
of the star Proxima Centauri in the pseudo-potential given by Eq (4.4) 
in \cite{popov, popov2},
\begin{equation} \label{Ups}
U_{ps} (\mathbf{r}) = \frac{G m M_S}{r_{SE}^2} \hat{\mathbf{r}}_{SE} \cdot
\mathbf{r} \,.
\end{equation}
Here $G$ stands for Newton's constant, $M_S$ stands for
the mass of the Sun and $\mathbf{r}_{SE}(t)$ describes the motion of the Sun in
the Earth's pseudo-potential and was calculated in \cite{popov}. 

The Lagrangian that describes the motion of the Proxima Centauri in the Earth's
pseudo-potential is therefore given by (gravitational interaction between the
star and the Sun is, of course, neglected):
\begin{equation} \label{L}
L = \frac{1}{2} m \dot{\mathbf{r}}^2 -
 \frac{G m M_S}{r_{SE}^2} \hat{\mathbf{r}}_{SE} \cdot \mathbf{r} \,,
\end{equation}
where $m$ is the mass of the star, and $\mathbf{r}(t)$ describes its motion. The
equations of motions are mass-independent, as expected.

\begin{figure}[t]
  \centering
  \includegraphics[scale=0.5]{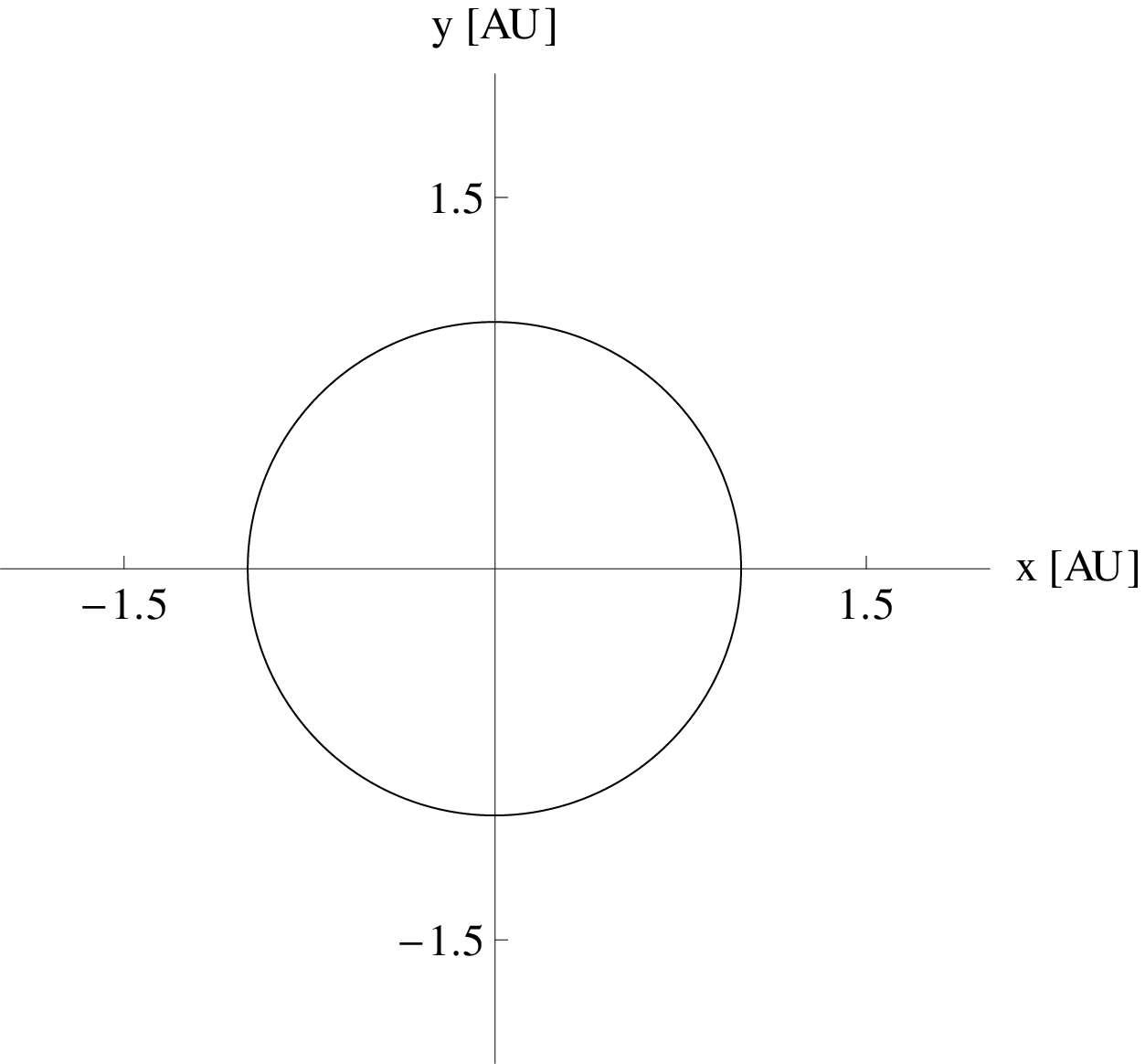} $\qquad$
  \includegraphics[scale=0.3]{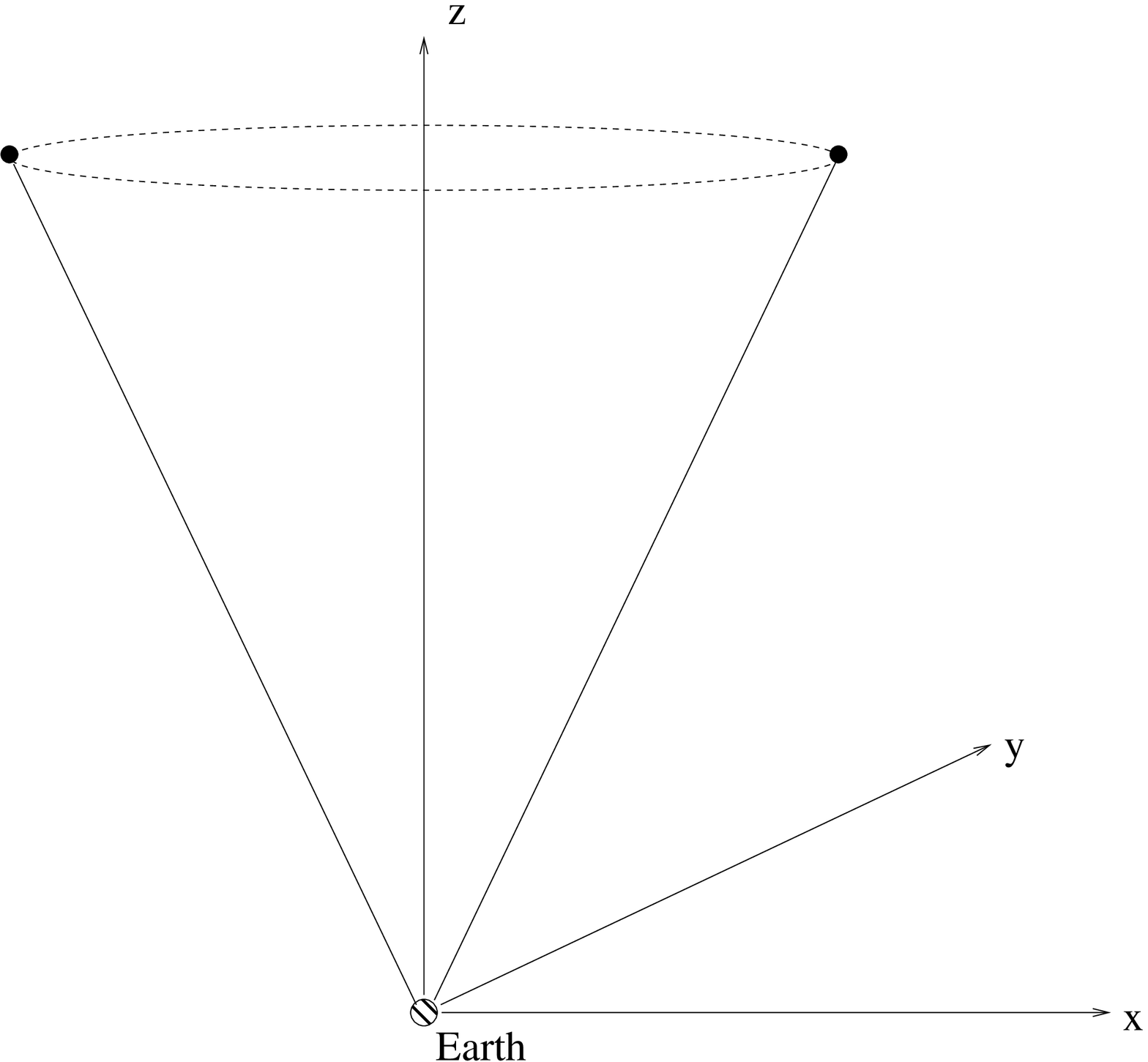}
  \caption{\label{fig2} Left pannel displays the result of the numerical
solutions for equations
    of motion derived from the Lagrangian (\ref{L}) over the period of 1 year.
It represents the trajectory of the star in the $x$-$y$ plane, as seen from the
Earth. Right pannel illustrates
    the stellar parallax effect, in consistence with the numerical results.}
\end{figure}

The Euler-Lagrange equations for this Lagrangian are solved numerically in the
Cartesian coordinate system, using \emph{Wolfram Mathematica} package. The
numerical solutions over the period of 1 year are presented in the Fig
\ref{fig2}. 

Stellar parallax can now be geometrically calculated:
\begin{equation} \label{parallax}
\arctan \theta = \frac{ r_x(t=0.5\textrm{ y}) }{ D },
\end{equation}
where $D=4.24$ ly is the well-knows distance of Proxima Centauri from the Earth
\cite{wiki}. Using the numerical results obtained above, one can evaluate the 
expression (\ref{parallax}). The result is
\begin{equation}
\theta = 3.705 \times 10^{-6}\textrm{ rad} = 0.76'' \,,
\end{equation}
which is perfectly consistent with the astronomical data \cite{mcarthur}.

\section{Conclusion}

We have analysed the motion of the star Proxima Centauri in the Earth's
pseudo-potential previously derived from Mach's principle \cite{popov}. The
obtained results are in accord with the observed data. The kinematical and
dynamical equivalence of Neo-Tychonian and Copernican systems has once again
been demonstrated.

{\ack Author kindly thanks Dr Karlo Lelas for bringing up this issue. This work
is supported by the Croatian Government under contract number 119-0982930-1016.}

\end{document}